\documentstyle[preprint,aps,psfig]{revtex}

\begin{document}

\draft
\preprint{ADP-94-6/T148}
\title{Pion Content of the Nucleon in Polarized Semi-Inclusive DIS
  \footnote{to appear in Proceedings of the Workshop on CEBAF
            at Higher Energies, CEBAF, April 1994}}
\author{W.Melnitchouk}
\address{Institute for Theoretical Physics,
         University of Regensburg,
         \mbox{D-93040} Regensburg, Germany.}
\author{A.W.Thomas}
\address{Department of Physics and Mathematical Physics,
         University of Adelaide,
         S.A. 5005, Australia.}

\maketitle

\begin{abstract}
An explicit pionic component of the nucleon
may be identified by measuring polarized
$\Delta^{++}$ fragments produced
in deep-inelastic scattering (DIS)
off polarized protons.
The pion-exchange model predicts highly correlated
polarizations of the $\Delta^{++}$ and
target proton, in marked contrast with the
competing diquark fragmentation process.
\end{abstract}

\section{Introduction}

Much has been learned about the internal structure of the nucleon
from inclusive deep-inelastic lepton scattering since the first
such experiments were carried out at SLAC in the late 1960s.
These and many subsequent DIS experiments [1]
were instrumental to the development of the quark-parton model
of hadrons.

Successful as it has been in describing much of the DIS data in
the perturbative region, a quark-parton based picture of the nucleon
is not entirely complete --- it is unable to explain
the non-perturbative structure of the nucleon.
A very good example of this is the deviation from the parton model
prediction for the Gottfried sum rule seen in the recent NMC
experiment [2].
The most natural explanation of this result is that there exists an
excess of $\bar d$ quarks over $\bar u$ in the proton, something
which is clearly impossible to obtain from perturbative QCD alone.
A non-perturbative pionic
cloud in the nucleon on the other hand offers a simple explanation
of SU(2) flavor symmetry breaking in the proton sea [3].

The pion content of the nucleon has been investigated in
previous studies of inclusive DIS [4],
although only upper bounds on the average pion number
were able to be extracted.
In addition, since the pion contribution to the nucleon structure
function appears at relatively small Bjorken $x$ ($x \sim 0.1$),
its signal is lost amongst the perturbative sea background.
It is therefore a challenge to seek direct experimental confirmation
of pionic effects in inclusive nucleon DIS.

The pertinent question is whether pions leave any unique
traces in other processes which cannot be understood in terms
of quarks and gluons alone.
A possible candidate for this is semi-inclusive DIS,
where a hadron is detected in the final state in
coincidence with the scattered electron.
We shall demonstrate that pions give rise to characteristic fragmentation
distributions in comparison with the predictions of parton model
hadronization.
These differences are significantly enhanced when
polarization degrees of freedom are considered.
We focus on semi-inclusive polarized $\Delta^{++}$
electroproduction from a polarized proton,
$e \vec p \rightarrow e' \vec\Delta^{++} X^-$.
(The decay products of $\Delta^+$ or $\Delta^0$ include
neutrals whose detection would be more difficult.)

Because the $g_1$ structure function of a pion is zero,
an unpolarized electron beam will suffice for this purpose.
The produced $\Delta$'s will appear predominantly in the target
fragmentation region, or backward hemisphere in the $\gamma^* p$
c.m. frame, and hence will appear slow in the laboratory frame.
With the high luminosity beam available at CEBAF the
rate of $\Delta^{++}$ production will generally be high.
Even though the efficiency with which low momentum $\Delta$'ss can be
accurately identified is lower than for fast baryons produced
in the forward
$\gamma^* p$ c.m. hemisphere, their detection may still be feasible
with the CEBAF Large Acceptance Spectrometer.

\section{Kinematics of Target Fragmentation}

We define our variables in the laboratory frame as follows:
$l, l'$ are the momentum four-vectors of the initial and final electrons,
$P_{\mu} = (M; 0, 0, 0)$ and
$p_{\mu} = (p_{0}; |{\bf p}| \sin\alpha \cos\phi,$
          $|{\bf p}| \sin\alpha \sin\phi,$
          $|{\bf p}| \cos\alpha)$
are the momentum vectors of the target proton and recoil $\Delta$,
respectively, and
$q_{\mu} = (\nu; 0, 0, \sqrt{\nu^{2} + Q^{2}})$ denotes the photon
4-momentum, defined to lie along the positive $z$-axis.
Then $\nu = E-E'$ is the energy transferred to the target,
$y = \nu / E = 1 - E'/E$ is the fractional energy transfer relative to the
incident energy, and $Q^2 = -q^2 = 2 M E x y$\ is minus the four-momentum
squared of the virtual photon, with
$x = Q^2 / 2 P \cdot q$.
With the CEBAF upgrade to $E \approx 8-10$ GeV, values of
$x \approx 0.13-0.14$ can be reached in the deep inelastic region
for $\nu \approx 8$ GeV and $Q^2 \approx 2$ GeV$^2$.

The four-momentum transfer squared between the proton and
$\Delta$ is
$t = (P-p)^2 = -p^2_T / \zeta\ +\ t_{max}$,\
which is bounded from above
by $t_{max} = -(M_{\Delta}^2 - M^2 \zeta) (1-\zeta)/\zeta$,
where $\zeta = p \cdot q / P \cdot q$\ \ is the light-cone fraction
of the target proton's momentum carried by the $\Delta$.
In terms of $t$, the three-momentum of the produced $\Delta$ is given by
\begin{eqnarray}
|{\bf p}| &=& \frac{1}{2M} \sqrt{(M^2 + M_{\Delta}^2 - t)^2
                                      - 4 M^2 M_{\Delta}^2}\ ,
\end{eqnarray}
so that in the lab. frame the slowest $\Delta$'s are those
for which $t \rightarrow 0$,
which occurs when $\zeta \rightarrow 1$.
As the upper limit on $\zeta$ is $1-x$, slow
$\Delta$ production also corresponds to the $x \rightarrow 0$ limit,
and the slowest possible particles produced at $\zeta = 1$
(at $x = 0$) have momentum
$|{\bf p}_{min}| = (M_{\Delta}^2 - M^2) / 2M \approx 340$ MeV.
For the pion-exchange process considered here, the peak in the
differential cross section occurs at $|{\bf p}| \sim 600$ MeV.
This corresponds to a total c.m. energy squared of the
photon--proton system of $W^2 = (P+q)^2 \sim 14$ GeV$^2$,
and a missing mass of $p_X^2 = (k+q)^2 \sim 0.8$ GeV$^2$.

In terms of the laboratory angle $\alpha$ between the $\Delta$
and $\gamma^*$ directions, where
\begin{eqnarray}
\cos\alpha
&=& \frac{ M_{\Delta}^{2} + (1-2 \zeta) M^{2} - t }
         { \sqrt{(M_{\Delta}^{2} - M^{2} - t)^{2} - 4 M^{2} t} },
\label{calpha}
\end{eqnarray}
production of $\Delta$'s will occur between $\alpha = 0$ and
$\alpha_{max}
= \cos^{-1} \left( \sqrt{1 - (M \zeta / M_{\Delta})^2}
            \right) \simeq 50^o$
for $\zeta \rightarrow 1$.

For studies of the spin dependence of the fragmentation process,
we require the target proton polarization to be
parallel to the photon direction, with the spin of the produced
$\Delta$ quantized along its direction of motion.
Experimentally, the polarization of the produced $\vec\Delta^{++}$
can be reconstructed from the angular distribution of its decay
products ($p$ and $\pi^+$).

\begin{figure}[h]
\centering{\ \psfig{figure=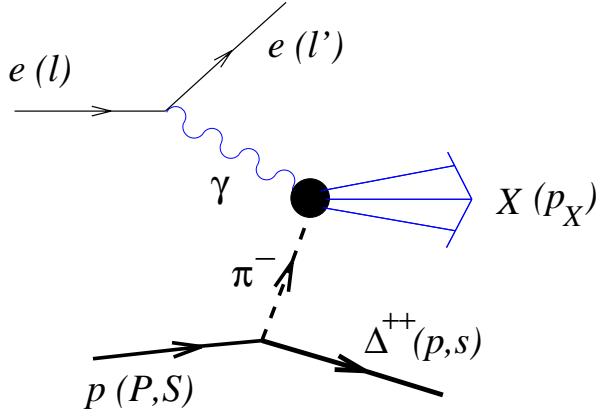,height=6cm}}
\caption{Semi-inclusive deep inelastic scattering of an electron from
   a polarized proton target, with a polarized $\Delta^{++}$ in the
   final state.}
\end{figure}

\section{Dynamics of Pion Exchange}

The relevant process is illustrated in Fig.1,
where the dissociation of a physical proton into a $\pi^-$ and $\Delta^{++}$
is explicitly witnessed by the probing photon.
In the pion-exchange model the differential cross section is
\begin{eqnarray}
{ d^5 \sigma \over dx dQ^2 d\zeta dp_T^2 d\phi }
&=& \left( { \alpha^2 \over M\ E^2\ x^2\ Q^4\ \zeta } \right)
    { f_{\pi N \Delta}^2 \over 16 \pi^2 m_{\pi}^2 }\
    { {\cal T}^{S\ s}(t)\ {\cal F}^2_{\pi \Delta} \over (t - m_{\pi}^2)^2 }\
    L_{\mu\nu}(l,q)\ W_{\pi}^{\mu\nu}(k,q)
\label{ope5}
\end{eqnarray}
where
$L_{\mu\nu} = 2 l'_{\mu} l_{\nu} + 2 l'_{\nu} l_{\mu} - g_{\mu\nu} Q^2$
is the lepton tensor,
and
\begin{eqnarray}
W_{\pi}^{\mu\nu}
&=& - \left( g^{\mu\nu} + \frac{ q^{\mu} q^{\nu} }{ Q^2 } \right)\
             W_{1\pi}\
 +\  \left( k^{\mu} + \frac{ k \cdot q }{ Q^2 } q^{\mu} \right)
     \left( k^{\nu} + \frac{ k \cdot q }{ Q^2 } q^{\nu} \right)\
     \frac{ W_{2\pi} }{ m_{\pi}^{2} }
\label{Wpi}
\end{eqnarray}
describes the $\gamma^* \pi$ vertex.
The pion structure function is extracted from
Drell-Yan experiments [5].

For the $\pi N \Delta$ form factor in Eq.(\ref{ope5})
we take the form suggested in Ref.[6]
in an analysis of the pionic content of the proton
in the infinite momentum frame,
\begin{eqnarray}
{\cal F}_{\pi \Delta}(p_T^2,\zeta) &=&
\left( { \Lambda^2 + M^2 \over
         \Lambda^2 + s_{\pi \Delta} }
\right)^2                               \label{6ffz}
\end{eqnarray}
where $s_{\pi \Delta} \equiv (m_{\pi}^2 + p_T^2)/(1-\zeta)
                           + (M_{\Delta}^2 + p_T^2)/\zeta$.
As discussed in Ref.[6], the formulation in the
infinite momentum frame allows the use of the experimental
structure function of the pion in Eq.(\ref{Wpi}).

The function ${\cal T}^{S\ s}(t)$ is obtained by
evaluating the trace over the target nucleon spinor and the
Rarita-Schwinger spinor-vector for the recoil $\Delta$.
Because it is emitted collinearly with the pion, production of
$\Delta$ baryons with spin projection $\pm 3/2$ is forbidden,
which leads to the selection rule:
\begin{eqnarray}
{\cal T}^{S\ \pm {3 \over 2}}(t) &=& 0.
\end{eqnarray}
The yield of spin projection $\pm 1/2$\ states is given by
\begin{eqnarray}
{\cal T}^{ +{1 \over 2}\ \pm{1 \over 2} }(t)
&=& { 1 \over 12 M_{\Delta}^2 }
    \left[ (M - M_{\Delta})^2 - t \right]\
    \left[ (M + M_{\Delta})^2 - t \right]^2\
    (1 \pm \cos\alpha)
\end{eqnarray}
where $\alpha$ is also the angle between the polarization
vectors $S$ and $s$.
Because the production of $\Delta$ baryons is limited to
forward laboratory angles, the presence of the $(1 \pm \cos\alpha)$ factor
associated with the final state polarization will significantly suppress
the $s = -1/2$ yield relative to that of $s = +1/2$ final states.
This suppression leads to strikingly different predictions for the
polarization asymmetry compared with those of the competing
parton fragmentation process discussed below.

\section{Backgrounds}

At the large $Q^2$ and $\nu$ possible with an 8--10 GeV electron
beam, the resonance backgrounds should not pose a major problem
in identifying the required signal.
Firstly, interference from quasi-elastic $\Delta^{++}$ production
will not be present because of charge conservation.
Secondly, the large $W (\sim 3-4$ GeV) involved means that
interference from excited $\Delta^*$ states (with
subsequent decay to $\Delta^{++}$ and pions)
will be negligible.
In addition, any such resonance contributions will be strongly
suppressed by electromagnetic form factors at large $Q^2$
($Q^2 \agt 2$ GeV$^2$).

A potentially more important background will be that due
to uncorrelated spectator fragmentation.
Within the parton model framework the cross section for this process
(assuming factorization of the $x$ and $\zeta$ dependence)
is proportional to [7]
$ {\cal F}_{p\uparrow}(x,Q^2)\
\widetilde{D}_{p\uparrow-q\uparrow\downarrow}^{s}(z,p_T^2)$,
where $z = \zeta/(1-x)$ is the light-cone momentum fraction
of the produced baryon carried by the spectator system.
The fragmentation function $\widetilde{D}(z,p_T^2)$ gives the
probability for the polarized
($p^{\uparrow}$ minus $q^{\uparrow\downarrow}$)
spectator system to fragment into a $\Delta^{++}$ with polarization $s$.
The usual assumption is that the transverse momentum distribution of the
$\Delta$ also factorizes,\
$\widetilde{D}_{p\uparrow - q\uparrow\downarrow}^s(z,p_T^2)
= D_{p\uparrow - q\uparrow\downarrow}^s(z)\ \varphi(p_T^2)$,
with $\int dp_T^2 \varphi(p_T^2) = 1$.

The function ${\cal F}_{p\uparrow}(x,Q^2)$ is proportional to the
spin-weighted interacting-quark momentum distribution functions,
$q^{\uparrow\downarrow} (x)$, where $^{\uparrow \downarrow}$ denote quark
spins parallel or antiparallel to the spin of the proton.
For estimation purposes we consider the model for
$q^{\uparrow\downarrow} (x)$ of Carlitz \& Kaur [8].
The results change little if one uses the model of Sch\"afer [9].

At large $z$ the fragment $\Delta^{++}$ carries most of
the parent system's momentum, and therefore contains
both valence $u$ quarks from the target proton.
In our region of interest ($z \agt 0.6$) by far the most important
contributions to $D(z)$ come from the process whereby the
$\Delta^{++}$ is formed after only one $u \bar{u}$ pair is created [10].
As a consequence DIS from valence $u$ quarks will be unimportant.
For scattering from sea quarks we assume the same fragmentation
probabilities for $uuq\bar q$ spectator states as for $uu$
(although at $Q^2 \simeq 2$ GeV$^2$ the sea constitutes at most
$\sim 15\%$ of the cross section at $x \sim 0.1$).
We thus parametrize the (very limited) EMC data [11]
on unpolarized $\Delta^{++}$ muon production
for $z \rightarrow 1$ as:
$D_{uu}(z \rightarrow 1) = a (1 - z)^b$,
where $a \approx 0.68$ and $b \approx 0.3$.

For the spin dependence of the fragmentation process
we follow the approach taken by Bartl et al. [12]
in their study of polarized quark $\rightarrow$ baryon fragmentation.
Namely, the diquark is assumed to retain its helicity
during its decay, and the $q \bar{q}$ pair creation probability
is independent of the helicity state of the quark $q$.
At leading order this means that the produced baryon contains
the helicity of the diquark, so that, for example,
a $\Delta^{\Uparrow}$ or $\Delta^{\uparrow}$ can
emerge from a $q^{\uparrow} q^{\uparrow}$ diquark,
whereas a $\Delta^{\Downarrow}$ cannot.
(Notation here is that $\Uparrow, \uparrow, \downarrow, \Downarrow$
represent $s = +3/2, +1/2, -1/2, -3/2$ states, respectively.)

The overall normalization of the spin-dependent fragmentation functions
is fixed by the condition
\begin{eqnarray}
         q(x)\ D_{p-q}(z)\
+\ \bar{q}(x)\ D_{p-\bar{q}}(z)\
&=& q^{\uparrow} (x)\
    D_{p\uparrow - q\uparrow}(z)\
 +\ q^{\downarrow} (x)\
    D_{p\uparrow - q\downarrow}(z)\       \nonumber\\
&+& \bar{q}^{\uparrow} (x)\
    D_{p\uparrow - \bar{q}\uparrow}(z)\
 +\ \bar{q}^{\downarrow} (x)\
    D_{p\uparrow - \bar{q}\downarrow}(z)  \label{norm}
\end{eqnarray}
where
$ D(z)
= \sum_{s=-3/2}^{+3/2} D^{s}(z)$.
In relating the production rates for various polarized $\Delta^{++}$
we employ SU(6) spin-flavor wavefunctions,
from which simple relations among the valence diquark $\rightarrow$
$\Delta^{++}$ fragmentation functions, $D_{ qq_{j(j_z)} }^{s}(z)$,
can be deduced (the diquark state $qq_{j(j_{z})}$ is labeled by
its spin $j$ and spin projection $j_z$).
The leading functions are related by:
\begin{eqnarray}
D_{ uu_{1(1)} }^{\Uparrow}(z)
= 3\ D_{ uu_{1(1)} }^{\uparrow}(z)
= \frac{3}{2} D_{ uu_{1(0)} }^{\uparrow}(z)
= \frac{3}{2} D_{ uu_{1(0)} }^{\downarrow}(z),  \label{leadFF}
\end{eqnarray}
with normalization determined from:
\begin{eqnarray}
D_{ uu_{1(1)} }^{\Uparrow}(z)
&=& \frac{3}{4} D_{ uu }(z).                    \label{polunp}
\end{eqnarray}
(Note that this is true only when the spin projections of the diquark
and $\Delta$ are in the same direction.)
The non-leading fragmentation functions are those which require
at least two $q \bar{q}$ pairs to be created from the vacuum, namely
$D_{ uu_{1(0)} }^{\Uparrow/\Downarrow}$,
$D_{ uu_{1(1)} }^{\downarrow/\Downarrow}$,
$D_{ ud_{0(0)} }^{\Uparrow/\uparrow/\downarrow/\Downarrow}$,
$D_{ ud_{1(0)} }^{\Uparrow/\uparrow/\downarrow/\Downarrow}$,
and
$D_{ ud_{1(1)} }^{\Uparrow/\uparrow/\downarrow}$,
and those which require 3 such pairs,
$D_{ uu_{1(1)} }^{\Downarrow}$ and
$D_{ ud_{1(1)} }^{\Downarrow}$.
Except at very small $z$ ($z \alt 0.2$) the latter functions
are consistent with zero [10].
For the 2-$q \bar{q}$ pair fragmentation functions, we also expect that
$ D_{ uu_{1(0)} }^{\Uparrow}(z)
= D_{ uu_{1(0)} }^{\Downarrow}(z)$.
For $z \agt 0.2$ the unpolarized model fragmentation functions of Ref.[10]
requiring two $q \bar{q}$ pairs (e.g. $D_{ud}(z)$) are quite
small compared with the leading fragmentation functions,
$D_{ud}(z) \simeq 0.1\ D_{uu}(z)$.
For spin-dependent fragmentation we therefore expect
a similar behavior for those decay probabilities requiring two
$q \bar{q}$ pairs created in order to form the final
state with the correct spin and flavor quantum numbers.
This then allows for a complete description of polarized fragmentation
at large $z$ in terms of only the 4 fragmentation functions in
Eq.(\ref{leadFF}).

Finally, the $p_T$-integrated differential cross section
for the electroproduction of a $\Delta^{++}$ with spin $s$
can be written:
\begin{eqnarray}
{ d^3\sigma \over dx dQ^2 d\zeta }
&=&
\left( { 2 \pi \alpha^2  \over  M^2 E^2 x (1-x) } \right)
\left( { 1 \over 2 x^2 }\
    +\ { 4 M^2 E^2 \over Q^4 }
       \left( 1 - {Q^2 \over 2 M E x} - { Q^2 \over 4 E^2 } \right)
\right)                                                      \label{qpmful}\\
& & \hspace*{-1.3cm} \times
\left[
{4 x \over 9}
\left( u_V^{\uparrow} D_{ud_{1(0)}}^{s}
     + 2 \bar{u}^{\uparrow} \left({2 \over 3} D_{uu_{1(1)}}^{s}
                           + {1 \over 3} D_{uu_{1(0)}}^{s}
                       \right)
     + u_V^{\downarrow} D_{ud_{1(1)}}^{s}
     + 2 \bar{u}^{\downarrow} \left({2 \over 3} D_{uu_{1(1)}}^{s}
                           + {1 \over 3} D_{uu_{1(0)}}^{s}
                       \right)
\right)
\right.                                                         \nonumber\\
& & \hspace*{-1.0cm} +
\left.
{x \over 9}
\left( d_V^{\uparrow} D_{uu_{1(0)}}^{s}
     + 2 \bar{d}^{\uparrow} \left({2 \over 3} D_{uu_{1(1)}}^{s}
                           + {1 \over 3} D_{uu_{1(0)}}^{s}
                       \right)
     + d_V^{\downarrow} D_{uu_{1(1)}}^{s}
     + 2 \bar{d}^{\downarrow} \left({2 \over 3} D_{uu_{1(1)}}^{s}
                           + {1 \over 3} D_{uu_{1(0)}}^{s}
                       \right)
\right)
\right].                                                        \nonumber
\end{eqnarray}

\section{Numerical Results and Discussion}

The differential cross section, $Q^2 d^3\sigma / dx dQ^2 d\zeta$,
for the individual polarization states of the produced $\Delta^{++}$
(for DIS from a proton with $S=+1/2$) is shown in
Fig.2, for $x = 0.14$, $Q^2 = 2$ GeV$^2$ and $E = 8$ GeV.
The pion-exchange model predictions (solid curves)
are for a form factor cut-off of $\Lambda = 800$ MeV,
which gives $< n >_{\pi\Delta}$ $\approx 0.02$ [6].
(For the same $< n >_{\pi\Delta}$ the cut-off in a $t$-dependent dipole
form factor would be $\sim 700$ MeV.)
The spectrum shows strong correlations between the polarizations
of the target proton and $\Delta^{++}$.
In the quark-parton model (dashed curves) the correlations
are significantly weaker, and the ratio of polarized $\Delta$'s
in this case is
$s = +3/2 : +1/2 : -1/2 : -3/2 \approx 3 : 2 : 1 : 0$.

\begin{figure}
\centering{\ \psfig{figure=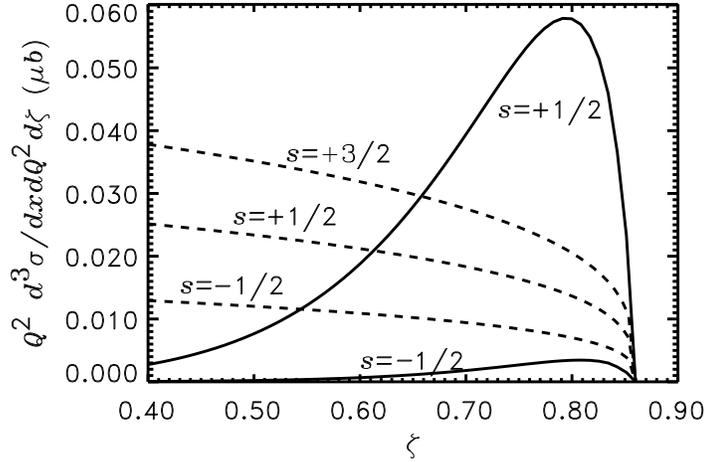,height=7cm}}
\caption{Differential electroproduction cross section for various
   polarization states of the $\Delta^{++}$.
   The $\pi$-exchange model, with cut-off mass $\Lambda=800$ MeV,
   is shown by solid curves.
   The quark-parton model background (dashed curves) is estimated using
   the fragmentation functions extracted from the unpolarized
   EMC data [11] and Eqs.(\protect\ref{leadFF})
   and (\protect\ref{polunp}).}
\end{figure}

The differences between the pion-exchange model and fragmentation
backgrounds can be further enhanced by
examining polarization asymmetries.
In Fig.3 we show the difference $\sigma^+ - \sigma^-$,
where $\sigma^{\pm} \equiv Q^2 d^3 / dx dQ^2 d\zeta (s=\pm 1/2)$,
as a fraction of the total unpolarized cross section.
The resulting $\zeta$ distributions are almost flat, but significantly
different for the two models.

\begin{figure}
\centering{\ \psfig{figure=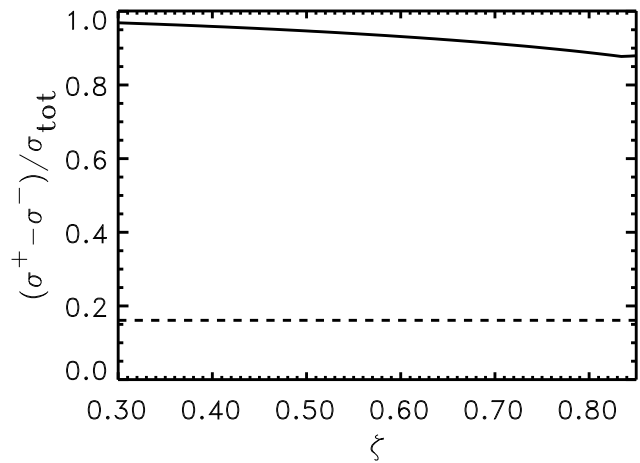,height=7cm}}
\caption{Polarization asymmetry for the $\pi$-exchange (solid)
   and parton fragmentation (dashed) models,
   with $\sigma^{\pm}$ as defined in the text,
   and $\sigma_{\rm tot}$ is the sum over all polarization states.}
\end{figure}

Of course the two curves in Fig.3 represent extreme cases,
in which $\Delta$'s are produced entirely via pion emission or
diquark fragmentation.
In reality we can expect a ratio of polarization cross sections
which is some average of the curves in Fig.3.
The amount of deviation from the parton model curve will
indicate the extent to which the pion-exchange process contributes.
{}From this, one can in turn deduce the strength of the $\pi N \Delta$
form factor.
Unlike inclusive DIS, which can only be used to place upper bounds
on the pion number, the semi-inclusive measurements could pin down
the absolute value of $< n >_{\pi\Delta}$.
A measurement of this ratio would thus be particularly useful in testing
the relevance of non-perturbative degrees of freedom in high energy
DIS processes.

\vspace*{0.5cm}
\parindent 0cm
{\bf References.}

[1]             E.D.Bloom, et.al.,
                {\em Phys.Rev.Lett.} {\bf 23} (1969) 930;
                M.Breidenbach, et.al.,
                {\em Phys.Rev.Lett.} {\bf 23} (1969) 935;
                J.J.Aubert et al. (EM Collaboration),
                {\em Phys.Lett.} {\bf 123B} (1983) 275.

[2]             P.Amaudruz et al. (New Muon Collaboration),
                {\em Phys.Rev.Lett.} {\bf 66} (1991) 2712.

[3]             E.M.Henley and G.A.Miller,
                {\em Phys.Lett.} {\bf B 251} (1990) 497;
                A.I.Signal, A.W.Schreiber and A.W.Thomas,
                {\em Mod.Phys.Lett.} {\bf A6} (1991) 271;
                W.Melnitchouk, A.W.Thomas and A.I.Signal,
                {\em Z.Phys.} {\bf A340} (1991) 85;
                S.Kumano and J.T.Londergan,
                {\em Phys.Rev.} {\bf D44} (1991) 717;
                A.Szczurek and J.Speth,
                {\em Nucl.Phys.} {\bf A555} (1993) 249.

[4]             J.D.Sullivan,
                {\em Phys.Rev.} {\bf D5} (1972) 1732;
                A.W.Thomas,
                {\em Phys.Lett.} {\bf 126B} (1983) 97.

[5]             B.Betev, et.al. (NA10 Collaboration),
                {\em Z.Phys.} {\bf C28} (1985) 15.

[6]             W.Melnitchouk and A.W.Thomas,
                {\em Phys.Rev.}~{\bf D~47}~(1993)~3794;
                A.W.Thomas and W.Melnitchouk,
                in: Proceedings of the JSPS-INS Spring School (Shimoda, Japan),
                (World Scientific, Singapore, 1993).

[7]             R.D.Field and R.P.Feynman,
                {\em Nucl.Phys.} {\bf B 136} (1978) 1;
                N.Schmitz,
                {\em Int.J.Mod.Phys.} {\bf A3} (1988) 1997;
                G.D.Bosveld, A.E.L.Dieperink and O.Scholten,
                {\em Phys.Lett.} {\bf B264} (1991) 11;
                W.Melnitchouk, A.W.Thomas and N.N.Nikolaev,
                {\em Z.Phys.} {\bf A342} (1992) 215.

[8]             R.Carlitz and J.Kaur,
                {\em Phys.Rev.Lett.} {\bf D 38} (1977) 674; 1102.

[9]             A.Sch\"afer,
                {\em Phys.Lett.} {\bf B 208} (1988) 175.

[10]            A.Bartl, H.Fraas and W.Majoretto,
                {\em Phys.Rev.} {\bf D 26} (1982) 1061.

[11]            M.Arneodo et al. (EM Collaboration),
                {\em Nucl.Phys.} {\bf B 264} (1986) 739.

[12]            A.Bartl, H.Fraas and W.Majoretto,
                {\em Z.Phys.} {\bf C 6} (1980) 335;
                J.F.Donoghue,
                {\em Phys.Rev.} {\bf D 17} (1978) 2922.

\end{document}